\documentclass{article}[11]

\usepackage{graphics,latexsym}
\usepackage{graphicx}
\usepackage{amsmath, amssymb,amsthm}

\usepackage{color}

\begin{document}

\title{Simulation of Electromagnetic Scattering \\ with Stationary
or Accelerating Targets}

\author{Daniele Funaro $^{1,*}$ and  Eugene Kashdan $^2$}
\date{}
\maketitle

\centerline{$^1$\small Dipartimento di Fisica, Informatica e Matematica }
\centerline{\small Universit\`a di Modena e Reggio Emilia, Via Campi 213/B, 41125
Modena (Italy)}
\medskip

\centerline{$^2$\small Numerical Solutions LLC }
\centerline{\small Tel Aviv, Israel}
\medskip

\centerline{\small $^*$Corresponding author: daniele.funaro@unimore.it}

\medskip

\begin{abstract} {\small
The scattering of electromagnetic waves by an obstacle is analyzed
through a set of partial differential equations combining the
Maxwell's model with the mechanics of fluids. Solitary type EM
waves, having compact support, may easily be modeled in this
context since they turn out to be explicit solutions. From the
numerical viewpoint, the interaction of these waves with a
material body is examined. Computations are carried out via a
parallel high-order finite-differences code. Due to the presence
of a gradient of pressure in the model equations, waves hitting
the obstacle may impart acceleration to it. Some explicative 2D
dynamical configurations are then studied, enabling the study of
photon-particle iterations through classical arguments.}
\end{abstract}

\vspace{.2cm}
\par\smallskip


%OCIS codes: 290.2558 Forward scattering, 290.5850  Scattering particles
% 000.4430  Numerical approximation and analysis
% 080.1753   Computation methods (geometrical optics)

\maketitle

\section{The model equations}\label{s:1}

The scope of this research is to investigate, from the numerical
viewpoint, physical phenomena arising as result of the scattering
of EM solitary waves with given obstacles. The model equations,
that include the classical Maxwell's system as a special case,
combine electromagnetism with inviscid fluid dynamics. As
introduced in \cite{fun}, by defining
$\rho=\vec\nabla\cdot\vec{E}$, the set of equations reads as
follows:
\begin{equation}\label{eq1}
\frac{\partial \vec{E}}
{\partial t} = c^2 \vec\nabla \times \vec{B}
-\rho\vec{V}\;, \quad \quad \quad\frac{\partial \vec{B}}{\partial t} = -\vec \nabla \times
\vec{E}\;, \quad \quad \quad\vec\nabla\cdot\vec{B}=0\;\\
\end{equation}
\begin{equation}\label{eq2}
\frac{\partial p}{\partial t}= \mu\rho \vec E\cdot \vec V\;\quad \quad\quad
\end{equation}
\begin{equation}\label{eq3}
\rho \biggl[\frac{D \vec{V}}{D t} +
\mu(\vec{E}+\vec{V}\times\vec{B})\biggr] =-\vec\nabla p\;
\end{equation}
where $\mu$ is a dimensional constant,  $D \vec{V}/Dt=\partial
\vec{V}/\partial t+(\vec{V}\cdot\vec\nabla)\vec{V}$ is the
substantial derivative, $p$ is a sort of pressure and $c$ is the
speed of light. The first equation is the Amp\`ere law and by
taking its divergence we get the continuity equation:
\begin{equation}\label{cont}
\frac{\partial\rho}{\partial t} = -\vec\nabla\cdot (\rho \vec V )
\end{equation}
Formula (\ref{eq2}) comes from energy preservation arguments.
Finally, the last equation coincides with the Euler equation for
the flow field $\vec V$, with a forcing term depending on the
electromagnetic field. Such a term recalls the Lorentz law for a
charged particle moving under the action of electric and magnetic
fields.
\smallskip

For a full discussion about the properties of the equations and
their solutions the reader is referred to \cite{fun}, where the
constant $\mu$ has been estimated to be of the order of $10^{11}$
Coulomb/Kilograms.  It has to be said that there is no contrast
with consolidated theories and the model equations conform to all
the usual physics requirements, such as: energy conservation,
existence of a Lagrangian, Lorentz invariance, covariance (see
\cite{fun2} and \cite{fun4}). For the sake of briefness, we do not
report here these results.
\smallskip

A similar set of equations is usually proposed in the study of
plasmas (see, e.g., \cite{jackson}, p.491), hence our choice looks
physically consistent. However, let us note that effective matter
is not involved here; thus the velocity field $\vec V$ is not
related to the movement of real particles, but to the transfer of
electromagnetic information from a point to another. Note in fact
that $\rho$ is not a mass density. For $\rho =0$, one gets the
standard set of Maxwell's equations in vacuum.
\smallskip

In a simplified version, the above model allows us to study EM
solitary waves not subjected to external perturbations ({\sl free
waves}). They are obtained in the special case when:
\begin{equation}\label{fw}
\vec{E}+\vec{V}\times\vec{B}=0 ~~~~~~~~~~p=0~~~~~~~~~~~~~~\frac{D
\vec V}{Dt}=0
\end{equation}
i.e., the stream-lines are straight lines. It is important to
observe that the divergence-free condition $\rho =0$ is not
imposed. As justified in \cite{fun} and \cite{fun5}, such a
feature is crucial to obtain EM solitary waves in vacuum, so
opening the path to a classical representation of photons as
self-contained packets carrying electromagnetic signals in the
standard way.
\smallskip

By integrating the divergence of $\vec E$ in the support $\Omega$
of a photon one gets: $\int_\Omega \rho =\int_{\partial\Omega}\vec
E \cdot \vec n =0$, where $\vec n$ is the normal external to the
boundary  $\partial\Omega$.  In this way, we can claim that there
is no violation of the Gauss condition.
\smallskip

Due to external reasons (as in the case when a wave-packet
encounters an obstacle), one may have: $D\vec V/Dt\ne 0$. Thus
$\vec V$ changes direction, the rays are curving, the
electromagnetic wave-fronts follow a new evolution and pressure
may develop. We can study in this way configurations obtained as a
result of the interactions of waves and material bodies. Due to
the increase of pressure according to equation (\ref{eq2}),
mechanical forces may act on the obstacle and force it to move.
This is what we would like to simulate with a series of numerical
experiments.

\section{Preliminary tests}\label{s:2}

The computational engine used for simulations is based on the
high-order accurate parallel finite-difference solver ERWIN (see
\cite{rep}), that has been developed at Tel Aviv University for
multi-dimensional systems of nonlinear time-dependent PDEs. The
solver is based on the 4-th order multi-step Runge-Kutta method
(RK4) for integration in time and the 4-th order accurate explicit
finite-difference scheme for spatial discretization. The spatial
parallelization is introduced through the MPI. The user can choose
between various types of boundary conditions in order to truncate
the computational domain (periodic, Neumann or Dirichlet). The
code was initially designed and optimized for the small (16 cores)
Linux cluster, but it has been successfully ported to the
super-computer environment. The results of numerical simulations
based on the code have been showcased in
 \cite{rk} and in the report \cite{cineca}.
\smallskip

To begin with, we are going to check the reliability of the code
on some exact solutions. In Cartesian coordinates $(x,y,z)$,
soliton-like EM waves solving the full set of equations
(\ref{eq1}), (\ref{eq2}), (\ref{eq3}) are obtained for instance as
follows:
\begin{equation}\label{sol}
\vec{E} = c\Big(0, f(y) g(x-ct), 0\Big)~~~~~ \vec{B} = \Big(0, 0,
f(y) g(x-ct)\Big)~~~~~  \vec{V} = (c, 0, 0)
\end{equation}
where the functions $f$ and $g$ are totally arbitrary. In
particular, if both have compact support the initial datum
$f(y)g(x)$ is contained in a box. Here the signal shifts along the
$x$-axis without diffusion at the speed of light as prescribed by
$\vec{V}$. The electric and magnetic fields stay orthogonal to the
direction of propagation. Since $\vec{E}+\vec{V}\times\vec{B}=0$,
we are in presence of a free-wave (see (\ref{fw})). The solution
in (\ref{sol}) is unbounded in the direction of the $z$-axis. It
is not difficult however to construct full solutions bounded in
all directions (see \cite{fun4}).
\smallskip

Note that Maxwell's equations (corresponding to $\rho =0$) do not
admit such a kind of solutions and this is the main reason why
photons cannot belong to ``classical'' physics. Including photons
in the solution space of a standard set of PDEs is the main
achievement of the model equations (\ref{eq1}), (\ref{eq2}),
(\ref{eq3}), with the aim to provide a classical background to
quantum theories. This important statement  is not developed in
this paper, where we are mainly concerned with some applications.
\smallskip

As expected, the experiments (not documented here) show the
correct shifting of the initial configuration along the direction
of $\vec V$. No significant diffusion is observed. Some
dissipation is introduced by the numerical scheme ({\sl artificial
viscosity}), which is however maintained within reasonable limits
and can be reduced by refining the discretization grids. Similar
experiments were performed in \cite{fun3}. There, the second-order
Lax-Wendroff scheme was used to study analogous situations and,
successively, to simulate the passage of a solitary EM wave
through an aperture.
\smallskip

\begin{figure}[!h]
\vspace{.1cm}
\centering
%\begin{centering}
%\vfill{}
\includegraphics[width=0.6\textwidth]{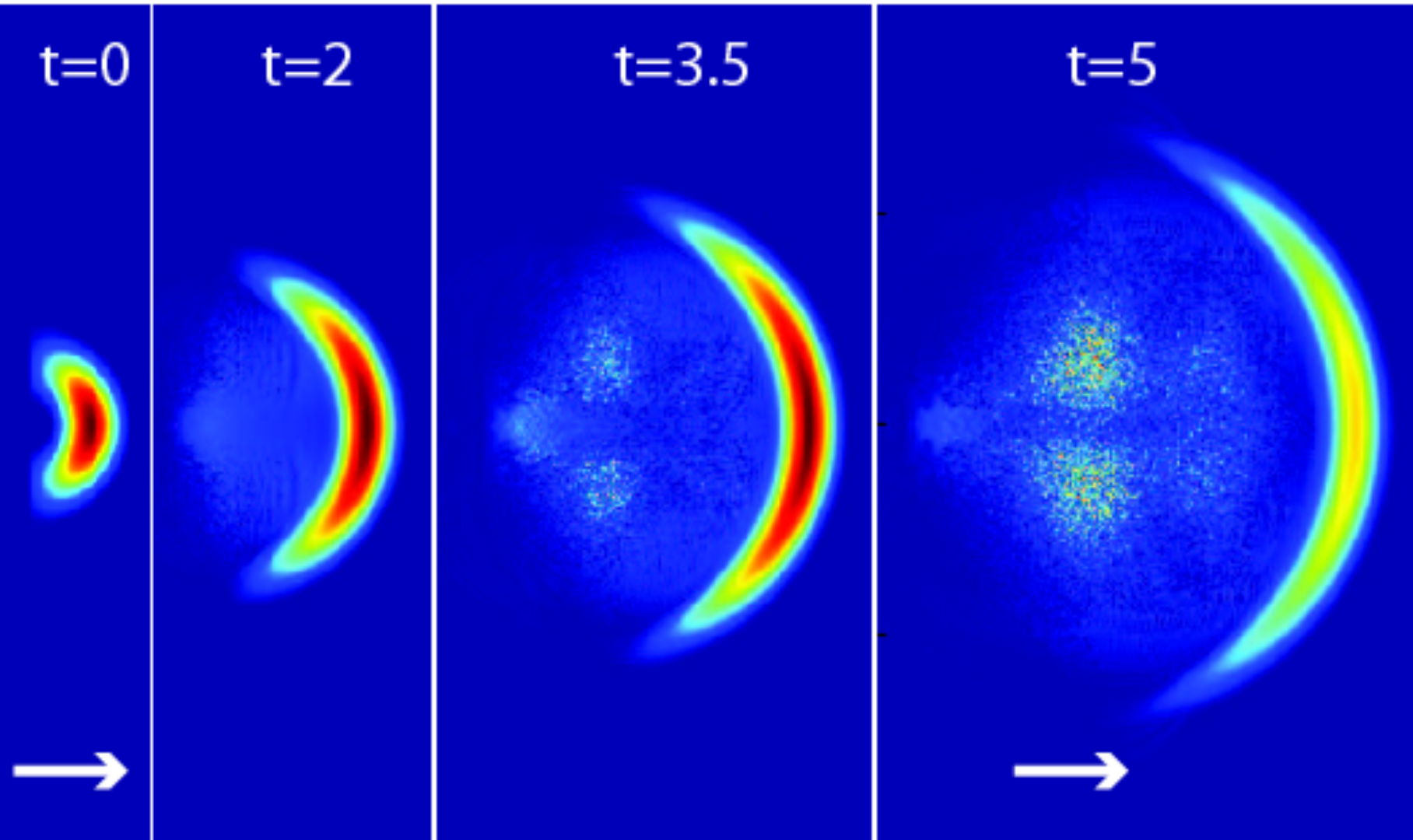}
%\hfill{}
\caption{\small \sl Intensity of the electric field at different
time snapshots for a banana-shaped pulse. Vector field $\vec E$
lays on the page (plane $(x,y)$) and is always orthogonal to the
direction of motion. The magnetic field is orthogonal to the page
($z$-axis). The rules of geometrical optics are fully preserved}
\label{banana}
\end{figure}

Remaining in the framework of the basic experiments, a more
advanced case is the one of Fig. \ref{banana}, where the initial
setting is such that the electric field $\vec E$ follows a
banana-like configuration, in such a way that the field $\vec V$
(orthogonal to both $\vec E$ and $\vec B$) is of radial type and
displays angles constrained in a fixed interval $[\theta_1
,\theta_2 ]$. At time $t=0$ we define (in polar coordinates $r$
and $\theta$):
\begin{align}\label{v0}
I_0=
\begin{cases}
\sin{\theta}~\sin^2\biggl[\frac{\pi(r-q)}{2p}\biggr],
~~&q<r<2p+q\;,~~\theta_1<\theta <\theta_2\;,\\
0, ~~&\text{otherwise}
\end{cases}
\end{align}
In transverse electric (TE) mode, the Cartesian coordinates of the
the nonzero initial conditions are defined as:
\begin{align*}
E_x=cI_0\sin\theta, && E_y=-cI_0\cos\theta\;, &&\text{~~and~~} ~~~~~~B_z=-I_0\;.
\end{align*}
As time evolves, the propagation speed $\vec{V}$ turns out to be:
\begin{align*}
 \vec{V}=c~ \frac{\vec{E}\times\vec{B}}{|\vec{E}\times\vec{B}|}
\end{align*}

In the numerical simulation shown in Fig. \ref{banana} we set
$p=1$ and $q=0.5$. In this and the subsequent simulations
showcased in this manuscript the values of time $t$ are given for
the qualitative purposes (to highlight the temporal evolution of
the modeled phenomena).
\smallskip

The fronts of this rounded pulse follows portions of concentric
circumferences. Condition $\vec E+\vec V\times\vec B=0$ is
satisfied at all times (see (\ref{fw})). The energy fades since it
is distributed on surfaces of increasing size. Some wiggles are
present because of minor inaccuracies of the discretization
method. As we already mentioned above such a wave-pulse has the
electric field transversal to the direction of motion, in perfect
agreement with the rules of geometrical optics (Huygens
principle). Note that, in such a circumstance, the finiteness of
the object requires the divergence $\rho$ to be different from
zero, especially near the furthermost tips. Therefore, the
patterns of Fig. \ref{banana} do not represent a part of a
classical Maxwellian cylindrical wave. Their behavior has chance
to be modeled only by the new set of equations.

\section{Scattering of solitons by a body}\label{s:3}

A fixed cylindrical obstacle, is now introduced in the path of the
wave. Since there is a deviation of the trajectories of the light
rays (i.e.: $D \vec{V}/Dt \not =0$), the full set of equations
must now be taken into account. We assume that the first two
components of $\vec{E}$ are different from zero, while the
magnetic field remains orthogonal to the plane $(x,y)$. In this
simplified case the polarization of the wave is basically fixed,
and we can study 2D configurations of the electric field. More
complex 3D problems could however be handled, though with an
increased degree of computational effort.
\smallskip

\begin{figure}[!h]
\vspace{.1cm}
\centering
%\begin{centering}
%\vfill{}
\includegraphics[width=0.45\textwidth]{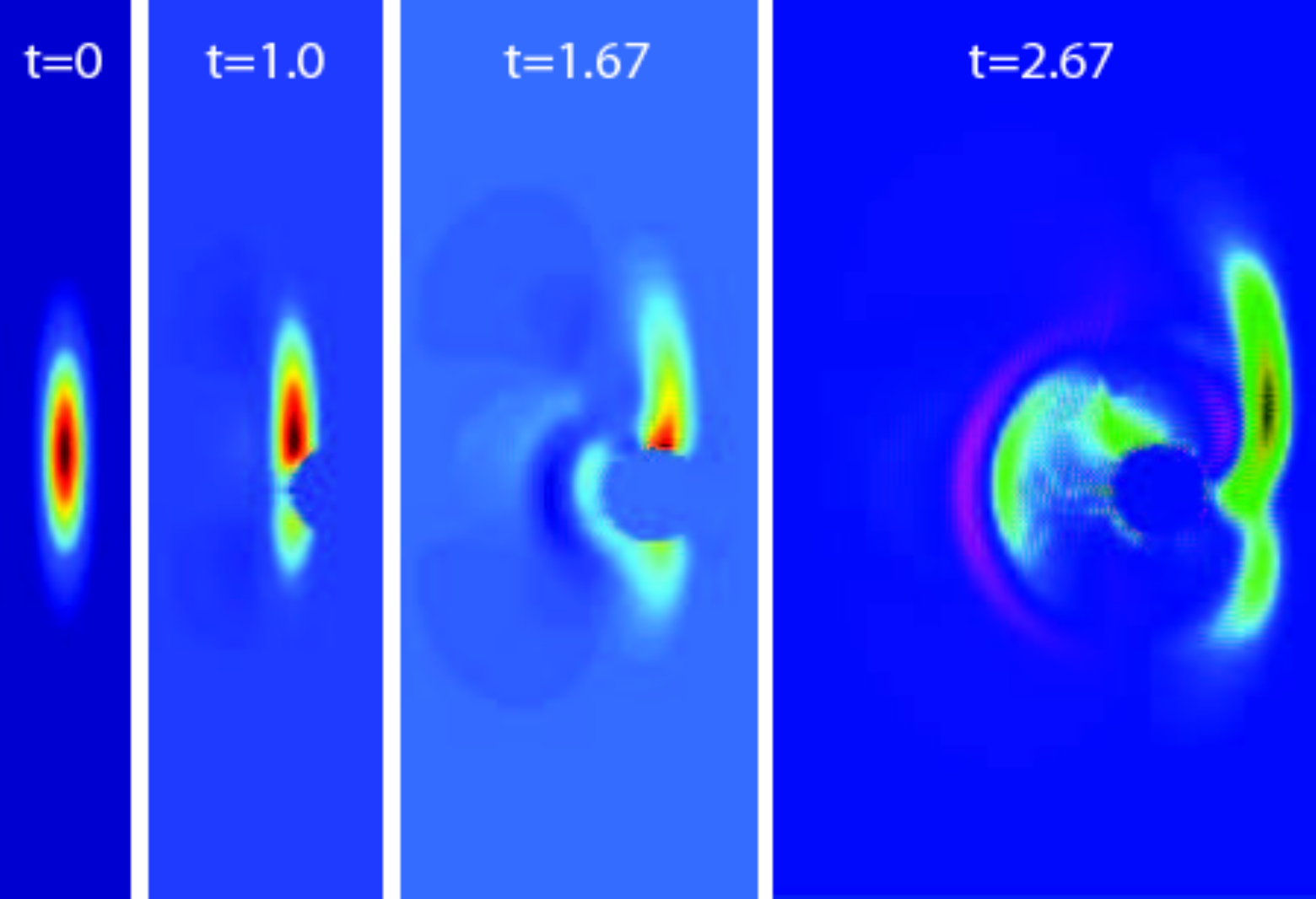}
%\hfill{}
\caption{\small \sl Encounter of an EM solitary
wave with a perfectly conducting cylindrical
obstacle.}\label{scat1}
\end{figure}

The results of a simulation (snapshots at different times) are
shown in Fig. \ref{scat1}. The initial condition is given by
$\vec{E}=c(0, f(y) g(x),0)$, where:
\begin{equation}\label{fg}
f(y)= \Big(1+\cos[\pi(y-y_0)/\sigma_1]\Big)^2~~~~~~~~~~~~
g(x)=\Big(1+\cos[\pi(x-x_0)/\sigma_2]\Big)^2
\end{equation}
Here  $\sigma_1$ and $\sigma_2$ are scaling constants  and
$(x_0,y_0)$ denotes the initial position. We set $\sigma_1=1$ and
$\sigma_2=6$ and center the wave along propagation axis. According
to (\ref{sol}), the successive free evolution is characterized by
a shifting along the $x$-axis at constant velocity $c$ without any
diffusion, until the obstacle is reached.
\smallskip

As the compact-support wave touches the body (a circular cylinder
of radius $r=0.5$ centered slightly below the horizontal symmetry
axis), the fields are modified based on the boundary conditions.
For instance, for a purely conductive body we must have that $\vec
E$ stays orthogonal to the obstacle's surface. This modifies the
balance $\vec E+\vec V\times\vec B =0$. The wave is not free
anymore and, following equation (\ref{eq3}), its trajectory
changes. This is what we actually imposed, by splitting normal and
tangential components on the obstacle boundary and only updating
the former. Nevertheless, with no practical impediment, other
boundary conditions could be taken into consideration, such as for
instance $\vec{E}=0$ or $\vec{V}=0$ (no slip condition).
\smallskip

The plots of Fig. \ref{scat1} show the evolution of the $E_y$
component of the electric field. For the initial data we set:
$\sigma_1=1$ and $\sigma_2=6$. Due to its position and size, in
this experiment the obstacle does not affect too much the path of
the soliton, though some scattering is observed.
\smallskip

\begin{figure}[!h]
\centering
\vspace{.1cm}
%\begin{centering}
%\vfill{}
\includegraphics[width=0.45\textwidth]{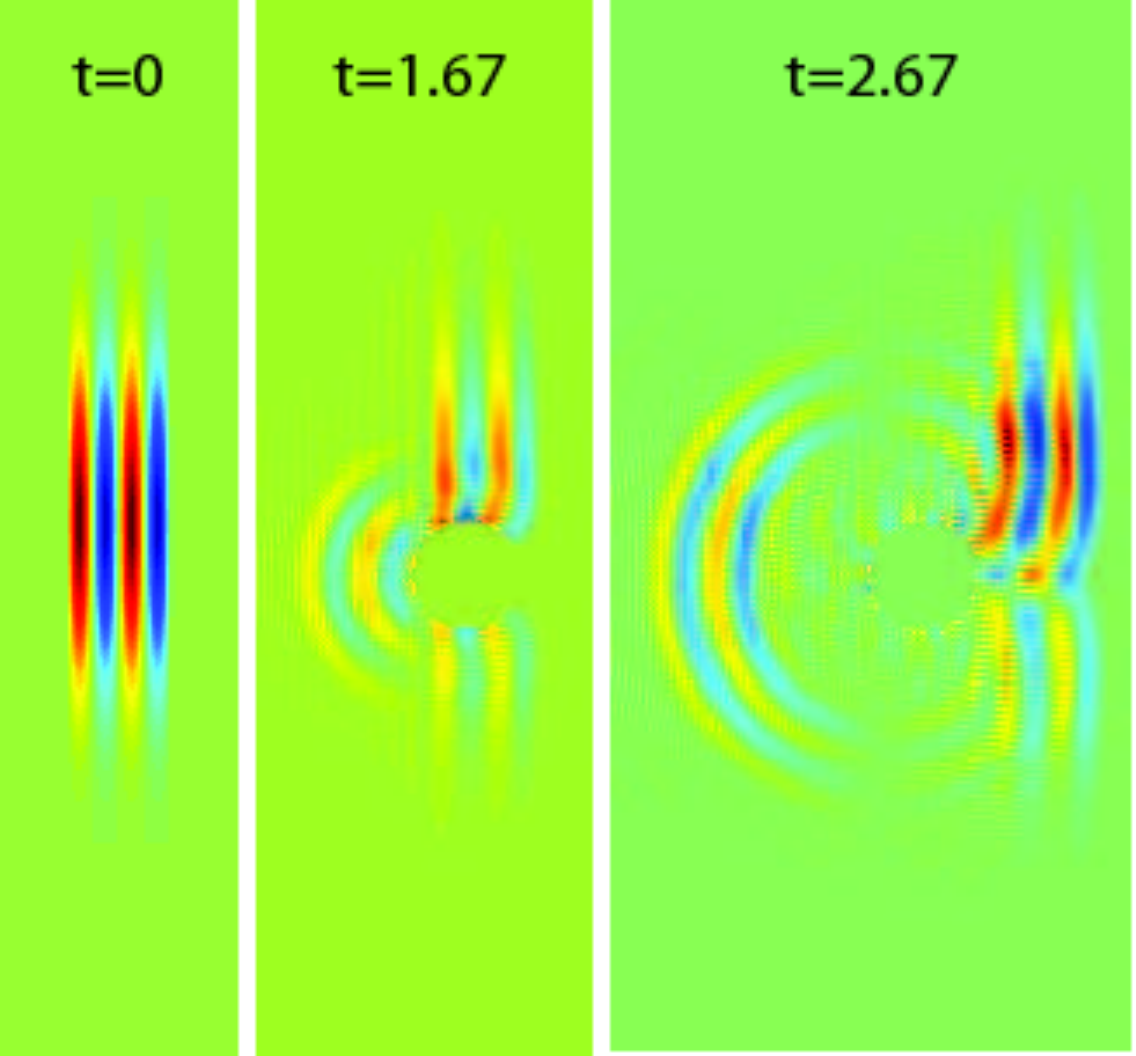}
%\hfill{}
\caption{\small \sl Scattering of a train of sine
waves by a perfectly conducting cylindrical
obstacle.}\label{scat2}
\end{figure}

In the successive experiment (Fig. \ref{scat2}), an incoming train
of sinus waves is obtained from (\ref{fg}) using the same scaling
parameters $\sigma_1=1$ and $\sigma_2=6$. As long as the wave
remains free there is no interference between the various pieces,
that independently shift at constant speed. Successively, the
radiation passes through the obstacle while part of it is
back-scattered. The results reasonably reproduce what is expected
from the real-life experiments.
\smallskip

Note that the waves shown in Fig. \ref{scat2} are not plane-waves
as it is usually assumed in this kind of experiments. Indeed, they
do not satisfy the condition $\rho =\vec{\nabla}\cdot \vec{E}=0$,
and thanks to this fact they may enjoy more freedom. Indeed, they
are allowed to stay bounded, which is a very realistic assumption.
Unbounded plane-waves carry infinite energy and are the only
possible solutions of (\ref{sol}) subject to the restriction $\rho
=0$ ($f$ is a constant). It is crucial to observe that Maxwell's
equations preserve the condition $\rho =0$ during the evolution
and this is why scattering experiments are usually initialized
with plane-waves. Regrettably, during and after the encounter with
the obstacle, $\rho$ may turn out to be different from zero;
nevertheless such a violation is rarely checked in numerical
tests. The modified equations can naturally handle the case $\rho
\not =0$, so we do not have to bother about imposing this
restriction and we can then concentrate out attention on a larger
range of boundary conditions. In other experiments (not shown
here) we considered the case where the wave also passes through
the obstacle having a different conductive constant (velocity less
than $c$).

\section{Accelerating the target}\label{s:4}

As a consequence of the use of equation (\ref{eq2}) a pressure
potential $p$ is developing in the neighborhood of the obstacle.
Through a multiplicative constant, the potential gradient can be
transformed into a force that pushes the body. As a result, the
body starts moving and the acceleration depends on the constant
$\mu$. We know that, in practice, light exerts pressure on matter,
though this effect is in general extremely mild. The situation is
enhanced when a photon hits a single massive particle. By choosing
appropriate values of $\mu$, we can more or less emphasize the
effect, depending on the scale we would like to simulate.
\smallskip

We show the results of the encounter  of a train of solitary
packets with a solid (non-deforming) cylindrical target similar to
one examined in the  previous section. However, in addition, we
model the accelerated movement of the target as a consequence of
the pressure gradient generated by the impact. For this purpose we
first compute the gradient of the pressure $\vec{\nabla}p$ (note
that in equation (\ref{eq2}) it is enough to break the
orthogonality of $\vec V$ and $\vec E$). Afterwards we compute the
acceleration $\vec{a}$ along the perimeter of the target by
applying the Second Newton's Law: $\vec{a}$=$\vec{\nabla}p/\mu$.
Since the body is considered rigid, the resultant is applied to
the barycenter. The variation of the target velocity is then
computed by $d\vec{v}=\vec{a}dt$, leading to a new displacement
$d\vec{s}$. At time $t=0$ we set the velocity equal to zero.
\smallskip

Depending on the wave-packet scaling parameters ($\sigma_1$,
$\sigma_2$) and the constant $\mu$, different scenarios may occur
in the  wave--target interaction. It has to be observed that both
$p$ and $\rho$ can assume positive or negative sign. This is a
novelty with respect to classical fluid dynamics, that allows for
the study of a more extended range of phenomena. Differences are
appreciated when varying initial location and size of obstacles,
as well as the polarization of the incoming wave-packet. Thus, the
speed and trajectory of the target depend on its size, its
location with respect to the wave propagation symmetry axis and
the dislocation of the scattered wave during the interaction
process.
\smallskip

\begin{figure}[!h]
\vspace{.1cm}
\centering
%\begin{centering}
%\vfill{}
\includegraphics[width=0.5\textwidth]{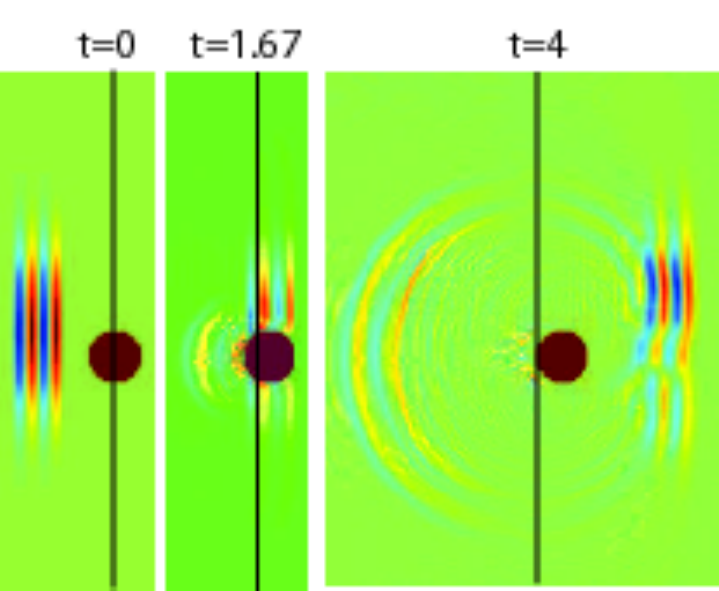}
%\hfill{}
\caption{\small \sl A train of solitary waves hits a relatively
large centered target imparting acceleration to it.}\label{scat3}
\end{figure}

We are ready to detail some possible behaviors. For the simulation
of Fig. \ref{scat3}, we have located a relatively large obstacle
($r=0.5$) below the longitudinal symmetry axis. A wave-packet,
consisting of two sinus periods, approaches the obstacle. The
first incoming front has a positive sign (corresponding to a
certain polarization). As the obstacle starts accelerating the
wave continues to pass producing more thrust until the two
entities are completely separated. In the figure, the fixed
vertical line indicates the location of the barycenter at initial
time $t=0$. As it should be, the speed $\vec v$ of the massive
obstacle turns out to be much lower than the speed of the
wave-packet $\vec V$ (proceeding at velocity $c$). A minor
vertical (transversal to the wave-packet direction)
oscillating-like movement is also observed.
\par\smallskip

\begin{figure}[!h]
\vspace{.2cm}
\centering
%\begin{centering}
%\vfill{}
\includegraphics[width=0.57\textwidth]{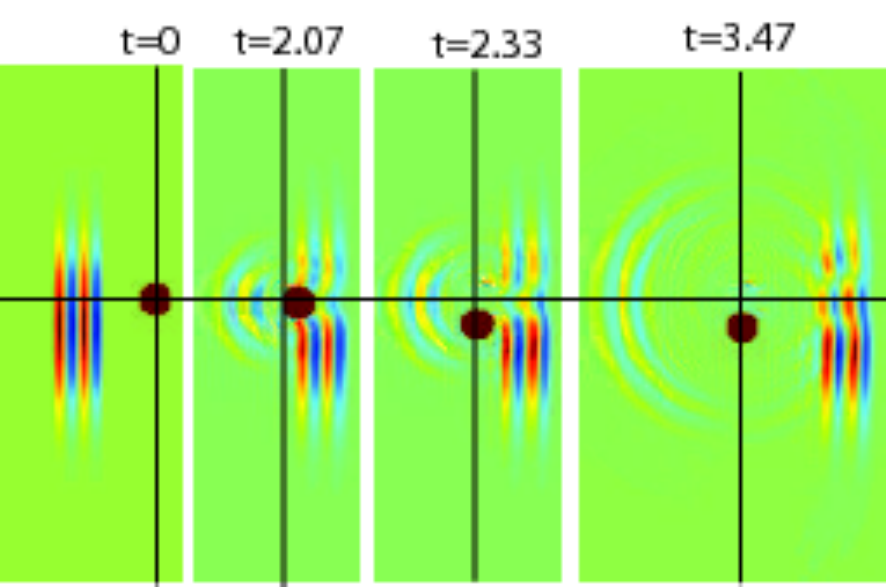}
%\hfill{}
\caption{\small \sl A train of solitary waves hits
a small target imparting acceleration to it.}\label{scat4}
\end{figure}

The results of a further simulation are shown in Fig. \ref{scat4}.
This time the sign of the incoming front has been switched
(reversed polarization). Moreover, the obstacle has a smaller size
($r=0.3$) and is located above the longitudinal symmetry axis.
Now, both backward and forward movements of the target have been
observed. The back-scattering also causes a significant shift in
the transversal direction. At the beginning, the target is pushed
forward. As it gets ``swallowed'' by the wave-packet, the
longitudinal component of the speed changes direction and the
transverse shift becomes dominant.
\smallskip

One should notice that in both simulations discussed here the
target movement is far from being smooth. For a little while, the
obstacle could follow a periodic orbit and suddenly move towards
another direction. This is particularly true for small targets.
Similar effects were observed in several other tests, not reported
here for the sake of brevity. We suspect that full 3D situations
can offer an even more variegated taxonomy.
\smallskip

At this point, it is worthwhile to make some remarks about the
application of computational techniques to EM problems.
Alternative versions of the Maxwell's model, similar to the one
here examined, have been proposed in order to stabilize numerical
schemes (see for instance \cite{taflove}) or to absorb outgoing
waves outside the computational domain (see \cite{abarbanel},
\cite{abarbanel2}, \cite{berenger}, \cite{yang}). The scope of
these modifications is not however to alter the nature of
Maxwell's equations as we are doing here, but  to solve numerical
troubles.
\smallskip

Consolidated finite-differences schemes (see, e.g., \cite{yee})
tend to mainly follow the evolution of the two dynamical equations
(Amp\`ere's  and Faraday's laws) and do not contemplate a
reinforcement of the divergence-free conditions. Very often,
Maxwell's equations are discretized through the so-called
$curl~curl$ formulation (well-suited in the framework of the
finite element method), where the divergence-free condition is not
actually imposed but appears as a penalty constraint. A viable
approach is to build approximation spaces satisfying
divergence-free conditions (\cite{assous}, \cite{bossavit},
\cite{cockburn}, \cite{hyman}, \cite{monk}). Divergence
corrections techniques can be also implemented (\cite{konrad},
\cite{munz}, \cite{rahman}). For a survey the reader is addressed
to \cite{schilders}. The reader can now understand why the removal
of condition $\rho=0$, compatibly with the physics of the problem
under study,  is so crucial especially for the construction of
numerical methods (see also the comments at the end of section 4).

\section{Conclusions}\label{s:5}

With the help of suitable model equations we were able to
numerically simulate the encounter of compact-support EM waves
with a solid obstacle. The EM pressure exerted can impart
acceleration to such an obstacle and the entire mechanism is still
described by the model with sufficient detail. It should be clear
that what we have done here may open the way to the quantitative
study of photon-particle scattering and the modeling of weak
interactions between particles. Such an analysis can be carried
out by directly following the laws of motion of the colliding
entities, and not by an inverse scattering procedure as is usually
done in these circumstances (see for instance \cite{ramm}). The
reliability of the results also depends on the choice of several
constants, that can be freely adjusted in setting up the numerical
tests.We have also shown that the back-scattering effect observed
experimentally plays significant role in accelerated particle
movement.
\smallskip

We also tried photon-photon scattering experiments,  but the
results require extra-care. The impact is very violent and
numerical instabilities may occur. On the other hand, the
introduction of artificial viscosity (for instance through the
addition of a second-order operator, such as the Laplacian, to the
Euler equation) introduces too much dissipation. The goal is to
improve the numerical scheme in order to come out with realistic
situations. This requires the adoption of discretization methods
for hyperbolic type equations which are well-suited to handle
shock waves.


\begin{thebibliography}{99}
\footnotesize

\bibitem{abarbanel} S. Abarbanel , D. Gottlieb, A mathematical analysis
of the PML method, J. Comput. Phys, 134 (1997), p. 357.

\bibitem{abarbanel2} S. Abarbanel, D. Gottlieb, J. S. Hesthaven,
Non-linear PML equations for time dependent electromagnetics in three
dimensions, J. of Sci. Comput., 28, 2-3 (2006), p. 125.

\bibitem{assous} F. Assous , P. Degond , E. Heintze , P. A. Raviart, J. Segre,
On a finite-element method for solving the three dimensional
Maxwell equations, J. Comput. Phys.,  109 (1993), p. 222.

\bibitem{berenger} J.-P. Berenger, A perfectly matched layer for the absorption
of electromagnetic waves, J. Comput. Phys., 114 (1994), p. 185.

\bibitem{bossavit} A. Bossavit, {\sl Computational Electromagnetism}, Academic
Press, Boston, 1998.

\bibitem{cockburn}  B. Cockburn, C.-W. Shu, Locally divergence-free discontinuous
Galerkin methods for the Maxwell equations, SIAM J. Numer. Anal.,
35 (1998), p. 2440.

\bibitem{fun} D. Funaro, {\sl Electromagnetism and the Structure of Matter},
World Scientific, Singapore, 2008.

\bibitem{fun2} D. Funaro, Electromagnetic radiations as a fluid flow,
arXiv:0911.4848v1

\bibitem{fun3} D. Funaro, Numerical simulation of electromagnetic solitons and their
interaction with matter, J. Sci. Comput., 45, 1 (2010), p. 259.

\bibitem{fun4} D. Funaro, A Lagrangian for electromagnetic solitary waves in vacuum,
arXiv:1008.2103v1

\bibitem{fun5} D. Funaro, From Photons to Atoms - The Electromagnetic Nature of Matter,
arXiv:1206.3110v1.

\bibitem{hyman}  J. M. Hyman, M. Shashkov, Natural discretization for the divergence,
gradient, and curl on logically rectangular grids, Computers Math.
Applic., 33-4 (1997), p. 81.

\bibitem{jackson} J.D. Jackson, {\sl Classical Electrodynamics}, Second Edition,
John Wiley \& Sons, New York, 1975.

\bibitem{rep} E. Kashdan, ERWIN -- The parallel high-order accurate solver for
multi-dimensional systems of nonlinear time-dependent PDEs with
the visualization engine, Tech. Report, School of Math Sciences,
Tel Aviv University, 2009.

\bibitem{cineca} E. Kashdan, D. Funaro, Interaction of electromagnetic
solitary waves, {\sl Science and Supercomputing in Europe: Research
Highlights 2010}, CINECA, Bologna, 2011, p. 138.

\bibitem{konrad}   A. Konrad, A method for rendering 3D finite element vector
field solution non-divergent, IEEE Trans. Magnetics, 25 (1989), p.
2822.

\bibitem{monk}  P. Monk, {\sl Finite Elements Methods for Maxwell's
Equations}, Oxford Univ. Press, New York, 2003.

\bibitem{munz}  C.-D. Munz, R. Schneider, E. Sonnendr\"ucker, U. Voss,
Maxwell's equations when the charge conservation is not satisfied,
C. R. Acad. Sci. Paris,  t.328, S\'erie I (1999), p. 431.

\bibitem{rahman} B.  Rahman, J. Davies, Penalty function improvement
of waveguide solution by finite elements, IEEE Trans. Microwave
Theory and Techniques, MTT-32 (1984), p. 922.

\bibitem{ramm} A.G. Ramm, {\sl Scattering by Obstacles}, Reidel, Dordrecht, 1986.

\bibitem{rk} P. Rosenau, E. Kashdan, Emergence of compact structures,
in a Klein-Gordon model, Phys. Rev. Lett., 104, 034101 (2010), and
$http://www.math.tau.ac.il/\sim compact$

\bibitem{schilders} W. H. A. Schilders, E. J. W. ter Maten  (Guest Editors),
{\sl Handbook of Numerical Analysis, Volume XIII, Numerical
Methods in Electromagnetics},  P. G. Ciarlet editor, Elsevier,
2005.

\bibitem{taflove} A. Taflove,  S. C. Hagness, {\sl Computational Electrodynamics:
The Finite-Difference Time-Domain Method}, Artech House, Norwood
MA, 2000.

\bibitem{yang} B. Yang, D. Gottlieb, J. S. Hesthaven, Spectral simulations
of electromagnetic wave scattering, J. Comput. Phys., 134-2
(1997), p. 216.

\bibitem{yee} K. Yee, Numerical solution of initial boundary value
problems involving Maxwell's equations in isotropic media, IEEE
Trans. on Antennas and Propagat., 14-3 (1966), p. 302.


\end{thebibliography}
\end{document}